\RequirePackage{lineno}
\documentclass[aps,prl,twocolumn,letterpaper,superscriptaddress]{revtex4}
\usepackage{graphicx}
\usepackage{amsmath,amssymb}
\usepackage{color}

\usepackage{latexsym,epsfig,multirow,comment,appendix,feyn,slashed,afterpage,makecell} 

\usepackage{mathrsfs}
\usepackage[x11names]{xcolor}
\usepackage{aas_macros}
\usepackage[normalem]{ulem}

\newcommand{\pd}{\partial}  
\newcounter{ichi}
\newcounter{ni}
\newcounter{san}
\newcounter{yon}
\begin{document}


\title{New prospects for detecting high-energy neutrinos from nearby supernovae}
\author{Kohta Murase}
\affiliation{Department of Physics; Department of Astronomy and Astrophysics; Center for Particle and Gravitational Astrophysics, The Pennsylvania State University, University Park, Pennsylvania 16802, USA}
\affiliation{Yukawa Institute for Theoretical Physics, Kyoto, Kyoto, 606-8502, Japan}

\begin{abstract}
Neutrinos from supernovae (SNe) are crucial probes of explosive phenomena at the deaths of massive stars and neutrino physics. High-energy neutrinos are produced through hadronic processes by cosmic rays, which are accelerated during interaction between the supernova (SN) ejecta and circumstellar material (CSM). Recent observations of extragalactic SNe have revealed that a dense CSM is commonly expelled by the progenitor star. We provide new quantitative predictions of time-dependent high-energy neutrino emission from diverse types of SNe. We show that IceCube and KM3Net can detect $\sim10^{3}$ events from a SN II-P (and $\sim3\times10^5$~events from a SN IIn) at a distance of 10~kpc. The new model also enables us to critically optimize the time window for dedicated searches for nearby SNe. 
A successful detection will give us a multienergy neutrino view of SN physics and new opportunities to study neutrino properties, as well as clues to the cosmic-ray origin. GeV-TeV neutrinos may also be seen by KM3Net, Hyper-Kamiokande, and PINGU.
\end{abstract}

\maketitle

\section{Introduction}
Thirty years ago, neutrinos from Supernova (SN) 1987A were detected by the Kamiokande-II~\cite{Hirata:1987hu} and Irvine-Michigan-Brookhaven~\cite{Bionta:1987qt} experiments. The neutrino detections confirmed that thermal neutrinos carry away the gravitational binding energy that is released in the core collapse~\cite{Colgate:1966ax}. While no neutrinos from Galactic supernovae (SNe) have been observed since the invention of the optical telescope and other multimessenger detectors, the core-collapse SN rate in the Milky Way is estimated to be $\sim3$ per century~\cite{Adams:2013ana}. If a Galactic SN occurs, high-statistics MeV neutrino signals will be seen by current facilities, enabling us to investigate details of core-collapse phenomena and neutrino oscillation in extreme environments.

Meanwhile, high-energy neutrino astrophysics has finally become a reality. High-energy cosmic neutrinos were discovered by the IceCube experiment~\cite{Aartsen:2013bka,Aartsen:2013jdh,Aartsen:2014gkd}. Nonthermal neutrinos are generated in the decay of charged pions produced by cosmic rays (CRs), via hadronuclear ($pp$) interactions with matter and photohadronic ($p\gamma$) interactions with radiation. They serve as a smoking gun of CR ion acceleration. No point source has been found yet, and the origin of the diffuse neutrino background is a big mystery in astroparticle physics~\cite{Murase:2015xka,Murase:2016gly,Halzen:2016gng}.   

\begin{figure}[th]
\includegraphics[width=\linewidth]{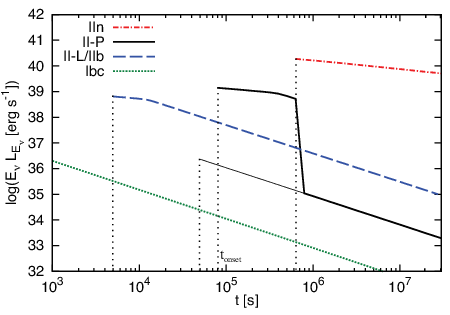}
\caption{Our predictions of neutrino ``light curves'' (at $E_\nu=1$~TeV) for various types of SNe. The slow decline implies the importance of late time emission. See text for details.
\label{fig1}
}
\end{figure}

SN remnants with an age of $\sim10^3-10^4$~yr are established as efficient particle accelerators~\cite{Funk:2015ena}. The theory also supports that CRs are accelerated at shocks via the Fermi mechanism, and it is believed that GeV-PeV CRs originate from SN remnants~\cite{Fermi:1949ee,Drury:1983zz,Caprioli:2015cda}. 
Are SNe (with an age of days to months) also promising CR and neutrino sources? Naively, the SN ejecta is freely expanding during the first $\sim1000$~yr, so the energy carried by CRs is so small that hadronic emission is difficult to detect (e.g.,~\cite{Chevalier:2016hzo,Berezhko:2015sma}). However, the situation has changed recently. Optical observations of various types of extragalactic SNe have provided cumulative evidence that a SN progenitor commonly experiences a significant mass loss a short time before the explosion~\cite{Smith:2014txa}. As a result, shock interactions with a dense circumstellar material (CSM) should occur, leading to efficient production of neutrinos. 

This work presents new time-dependent calculations of high-energy neutrino emission from nearby SNe with dense CSM interactions, and for the first time we provide detailed high-energy neutrino light curves from different classes of SNe (see Fig.~\ref{fig1}). The results, taking account of both time and energy dependence, are crucial to evaluate the signal-to-background ratio and examine the detectability with current and future detectors. We show that, $\sim0.1-10$~days after detections of MeV neutrinos and gravitational waves, a high-statistics TeV neutrino signal in IceCube and KM3Net is expected even for an ordinary Galactic SN. Our results suggest that nearby SNe may provide the first example of multienergy neutrino view of astrophysical objects. 

\begin{table}[t]
\begin{center}
\caption{CSM parameters for various types of SNe considered in this work. For SNe IIn and SNe II-P with an enhanced CSM, we also assume that the CSM is extended to $R_w=10^{16}$~cm~\cite{Yaron:2017umb} (implying $M_{\rm cs}\sim3~M_\odot$) and $R_w=4\times{10}^{14}$~cm~\cite{Ofek:2013afa} (implying $M_{\rm cs}\sim{10}^{-3}~M_\odot$), respectively.    
\label{tb1}
}
\scalebox{0.95}{
\begin{tabular}{|c||c|c|c||c|}
\hline Class & $D_*$ & $\dot{M}_{w}$ [$M_{\odot}$~yr$^{-1}$] & $V_{w}$ [km~s$^{-1}$] & $R_*$ [cm]\\
\hline IIn & $1$ & ${10}^{-1}$ &  $100$ & ${10}^{13}$\\
\hline II-P\footnote{With an enhanced CSM, based on SN 2013fs (II-P).} & ${10}^{-2}$ & ${10}^{-3}$ & $100$ & $6\times{10}^{13}$\\
\hline II-P\footnote{Without an enhanced CSM, based on Betelgeuse (RSG).} & $1.34\times{10}^{-4}$ & $2\times{10}^{-6}$ & $15$ & $6\times{10}^{13}$\\
\hline II-L/IIb & ${10}^{-3}$ & $3\times{10}^{-5}$ &  $30$ & $6\times{10}^{12}$\\
\hline Ibc & ${10}^{-5}$ & ${10}^{-5}$ & $1000$ & $3\times{10}^{11}$\\
\hline
\end{tabular}
}
\end{center}
\end{table}

\section{CSM interaction and CR acceleration}
We consider a SN explosion with a kinetic energy of ${\mathcal E}_{\rm ej}=10^{51}~{\rm erg}~{\mathcal E}_{\rm ej,51}$. After the shock breakout from a progenitor star, the SN ejecta with an ejecta mass of $M_{\rm ej}=10~M_\odot M_{\rm ej,1}$ starts to interact with a CSM (that is also used for an extended envelope in this work) with a density profile of $\varrho_{\rm cs}=Dr^{-w}$. We adopt a wind profile with $w=2$, which is reasonable in many cases. In the wind case, $D\equiv5\times{10}^{16}~{\rm g}~{\rm cm}^{-1}~D_*$ is related to the mass-loss rate ($\dot{M}_{w}$) and wind velocity $V_w$ as $D=\dot{M}_{w}/(4\pi V_w)$. It is noteworthy that recent observations have revealed that significant mass ejections or envelope inflations are common $\sim0.1-10$~yr before the core-collapse~(e.g., Refs.~\cite{Smith:2014txa,Smith:2007cb,Immler:2007mk,Miller:2008jy,Ofek:2013mea,Margutti:2013pfa,Margutti:2016wyh}), including the ``dominant''  SN class, SNe II-P~\cite{Morozova:2016efp,Yaron:2017umb}. For example, early observations of SN 2013fs indicated $D_*\sim{10}^{-2}$ and an outer edge radius of $R_w\sim{\rm a~few}\times10^{14}$~cm~\cite{Yaron:2017umb}. The most extreme class is Type IIn SNe~\cite{Ofek:2013afa,Fransson:2013qya,Ofek:2014fua}, and SN 2010jl inferred $D_*\sim6$ and $R_w\sim10^{16}$~cm~\cite{Ofek:2013afa}. A dense CSM is suggested in even Type Ibc SNe and low-luminosity $\gamma$-ray bursts~\cite{Hosseinzadeh:2016ysu,Campana:2006qe}. See Ref.~\cite{Smith:2014txa} and Table~\ref{tb1}.

A faster component of the SN ejecta is decelerated earlier, and the shock evolution is given by self-similar solutions~\cite{Chevalier:1982,Nadezhin:1985,Moriya:2013hka}. 
For an outer ejecta profile of $\varrho_{\rm ej}\propto t^{-3}{(r/t)}^{-\delta}$, the shock radius is given by~\cite{Chevalier:1982,Nadezhin:1985,Moriya:2013hka} 
\begin{equation}
R_{s}=X(w,\delta)D^{-\frac{1}{\delta-w}}{\mathcal E}_{\rm ej}^{\frac{\delta-3}{2(\delta-w)}}{M}_{\rm ej}^{-\frac{\delta-5}{2(\delta-w)}}t^{\frac{\delta-3}{\delta-w}},
\label{Rs}
\end{equation}
where $X(w,\delta)={[(3-w)(4-w)]}^{\frac{1}{\delta-w}}{[10(\delta-5)]}^{\frac{\delta-3}{2(\delta-w)}}$\\${[4\pi(\delta-4)(\delta-3)\delta]}^{-\frac{1}{\delta-w}}{[3(\delta-3)]}^{-\frac{\delta-5}{2(\delta-w)}}$ for the flat core profile. The solutions remain valid until the whole ejecta starts to be decelerated~\footnote{This occurs when the shock radius reaches $R_{\rm sh}=V_t t_t$, where $V_t={[10(\delta-5){\mathcal E}_{\rm ej}/3/(\delta-3)/M_{\rm ej}]}^{1/2}$~\cite{Chevalier:1982,Nadezhin:1985,Moriya:2013hka}.}, which is satisfied in our setup.
Progenitors of Type II-P SNe are thought to be red supergiants (RSGs), for which we assume a stellar size of $R_*=6\times{10}^{13}$~cm. For SNe II-L/IIb, we use a value motivated by yellow supergiants~\cite{Smith:2014txa}. We adopt $\delta=12$ for supergiant stars with a convective envelope, while $\delta=10$ is assumed for Wolf-Rayet-like compact stars with a radiative envelope~\cite{Matzner:1998mg}. For SNe IIn, we simply take $\delta=10$ based on the results on SN 2010jl~\cite{Ofek:2013afa}. 

While we use Eq.~(\ref{Rs}) for numerical calculations, for the demonstration we give expressions using Type II-P SNe as a reference. The shock radius is estimated to be
\begin{equation}
R_{s}\approx2.4\times{10}^{14}~{\rm cm}~D_{*,-2}^{-1/10}{\mathcal E}_{\rm ej,51}^{9/20}M_{\rm ej,1}^{-7/20}t_{5.5}^{9/10}
\label{RsIIP}
\end{equation}
and the corresponding shock velocity $V_s=dR_s/dt$ is: 
\begin{equation}
V_{s}\approx6.2\times{10}^{8}~{\rm cm}~{\rm s}^{-1}~D_{*,-2}^{-1/10}{\mathcal E}_{\rm ej,51}^{9/20}M_{\rm ej,1}^{-7/20}t_{5.5}^{-1/10}.
\label{VsIIP}
\end{equation}

Shock dissipation converts the kinetic energy into heat, magnetic fields, and CRs. The kinetic luminosity, $L_s=2\pi \varrho_{\rm cs}V_s^3R_s^2$ is estimated to be
\begin{equation}
L_s\approx1.0\times{10}^{42}~{\rm erg}~{\rm s}^{-1}~D_{*,-2}^{7/10}{\mathcal E}_{\rm ej,51}^{27/20}M_{\rm ej,1}^{-21/20}t_{5.5}^{-3/10}.
\label{LdIIP}
\end{equation}
Note that higher-velocity ejecta are more efficiently dissipated (the dissipation energy is given by ${\mathcal E}_{\rm ej}(>V)\propto V^{5-\delta}$), so the neutrino detectability is significantly enhanced compared to that in the simplest model with a uniform velocity~\cite{Murase:2010cu,Petropoulou:2016zar}. Also, unlike SNe IIn~\cite{Murase:2010cu,Petropoulou:2016zar}, the validity of the self-similar solution and the CR acceleration is justified in ordinary SNe II-P and II-L/IIb.

By analogy with SN remnants, it is natural to expect that CRs are accelerated by the shock acceleration mechanism. Contrary to the SN shock inside a star~\footnote{Since the stellar envelope is dense has a very steep profile, the radiative acceleration is relevant~\cite{Katz:2011zx}. The formation of collisionless shocks associated with the shock breakout from a star is not guaranteed.}, the CSM is not too dense (except for SNe IIn~\cite{Murase:2010cu,Murase:2013kda}) and the formation of collisionless shocks (mediated by plasma instabilities) is guaranteed. 
The condition for the shock to be radiation unmediated coincides with that for photons to breakout from the CSM~\citep{Murase:2013kda,Katz:2011zx,Kashiyama:2012zn}, which is $t\geq t_{\rm bo}\approx6.0\times{10}^3~{\rm s}~D_{*,-2}\mu_e^{-1}$ (where $t_{\rm bo}$ is the photon breakout time~\cite{Chevalier:2011ha}). 
In addition, since we consider CR acceleration during CSM interactions, we take the second criterion, $t\geq t_*\approx6.8\times{10}^{4}~~{\rm s}~D_{*.-2}^{1/9}{\mathcal E}_{\rm ej,51}^{-1/2}M_{\rm ej,1}^{7/18}R_{*,13.78}^{10/9}$, which is given by $R_*=R_s(t_*)$ for $V_s<V_{s,\rm max}$ (where $V_{s,\rm max}$ is the maximum velocity~\cite{Matzner:1998mg}). Considering these, the ``onset'' time of CR acceleration is given by 
\begin{equation}
t_{\rm onset}\approx {\rm max}[t_{\rm bo},t_*].
\end{equation}
We find that in most cases including dominant Type II-P SNe, $t_{\rm onset}\sim t_*$, which is ``different'' from $t_{\rm onset}\sim t_{\rm bo}$ for Type IIn SNe. See Fig.~\ref{fig1} for $t_{\rm onset}$ of various SN classes. 

The CR acceleration time is estimated to be $t_{\rm acc}\approx(20/3)cE_p/(eBV_s^2)$~\cite{Drury:1983zz}.  
In most cases listed in Table~\ref{tb1}, the maximum proton energy ($E_p^M$) is limited by the particle escape or dynamical time ($t_{\rm dyn}\approx R_s/V_s$)~\cite{Murase:2013kda}.
We assume a CR spectrum to be $dN_{\rm cr}/dp \propto p^{-s}$ with $s\sim2.0-2.2$, where $p$ is the proton momentum. The CR luminosity $L_{\rm cr}$ is normalized as $L_{\rm cr}=\epsilon_{\rm cr}L_s$, where $\epsilon_{\rm cr}\sim0.1$ is the energy fraction carried by CRs~\cite{Caprioli:2013dca}. 

\begin{figure}[tb]
\includegraphics[width=\linewidth]{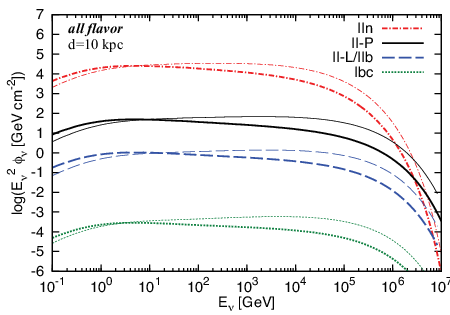}
\caption{Energy fluences of $\nu_e+\bar{\nu}_e+\nu_\mu+\bar{\nu}_\mu+\nu_\tau+\bar{\nu}_\tau$ from a Galactic SN at $t_{\rm max}$ for $d=10$~kpc (see~Table~\ref{tb2}). Thick and thin curves represent $s=2.2$ and $s=2.0$, respectively. 
\label{fig2}
}
\end{figure}

\section{Neutrino Production}
High-energy CRs interact with cold CSM nucleons and produce mesons (mostly pions) via inelastic $pp$ collisions, and high-energy neutrinos are generated via decay processes like $\pi^+\to\mu^+\nu_\mu\to\nu_\mu\bar{\nu}_\mu\nu_ee^+$. For a CR proton with $E_p$, the typical neutrino energy is $E_\nu\sim(0.03-0.05)E_p$~\cite{Kelner:2006tc}. The approximate cross section and proton inelasticity are $\sigma_{pp}\approx3\times{10}^{-26}~{\rm cm}^2$ and $\kappa_{pp}\approx0.5$, respectively. Using Eqs.~(\ref{RsIIP}) and (\ref{VsIIP}), the effective $pp$ optical depth is estimated to be
\begin{eqnarray}
f_{pp}&\approx&\kappa_{pp}\sigma_{pp}(\varrho_{\rm cs}/m_H)R_{s}(c/V_s)\nonumber\\
&\simeq&0.82~D_{*,-2}^{6/5}M_{\rm ej,1}^{7/10}{\mathcal E}_{\rm ej,51}^{-9/10}t_{5.5}^{-4/5}.
\label{fppIIP}
\end{eqnarray}

We numerical calculate neutrino spectra (see Supplemental Material for details), considering the detailed $pp$ cross section and the secondary spectra~\cite{Kelner:2006tc,Kafexhiu:2014cua}. For a given shock evolution with parameters listed in Table~\ref{tb1}, we evaluate $E_p^M$ and obtain time-dependent secondary spectra by solving kinetic equations.

In Fig.~\ref{fig1}, we first show TeV neutrino light curves for various types of SNe. A core-collapse SN event with a MeV neutrino luminosity of $L_\nu\sim{\rm a ~few}\times{10}^{53}~{\rm erg}~{\rm s}^{-1}$ will be accompanied by high-energy neutrino emission with a bolometric luminosity of $L_\nu \sim10^{37}-{10}^{42}~{\rm erg}~{\rm s}^{-1}$. For SNe II-P, the thick (thin) curve represents the case with (without) an enhanced CSM. The neutrino luminosity is expressed as $E_\nu L_{E_\nu}\propto \epsilon_{\rm cr}{\rm min}[1,f_{pp}]L_s$, so the slope changes at the time when $f_{pp}$ becomes unity. This can be seen for SNe II-P at $\sim4\times10^5$~s and SNe II-L/IIb at $\sim10^4$~s, respectively. For these types of SNe (with $D_*\sim10^{-4}-{10}^{-2}$), the best time window for high-energy neutrino observations is $\sim0.1-1$~day. On the other hand, for Type IIn SNe (with $D_*\sim10^{-1}-{10}$), $\sim0.1-1$~yr observations are necessary.
Note that the appropriate time window depends on SN types and the time-dependent model is critical.

Using Eqs.~(\ref{fppIIP}) and (\ref{LdIIP}), the neutrino energy fluence (for the sum of all flavors) is estimated to be
\begin{eqnarray}
E_\nu^2 \phi_{\nu}&\sim&\frac{1}{4\pi d^2}\frac{1}{2}{\rm min}[1,f_{pp}]\frac{\epsilon_{\rm cr}{\mathcal E}_{s}}{{\mathcal R}_{\rm cr10}}{\left(\frac{E_\nu}{0.4~{\rm GeV}}\right)}^{2-s}\nonumber\\
&\simeq&83~{\rm GeV}~{\rm cm}^{-2}~{\rm min}[1,f_{pp}]\epsilon_{\rm cr,-1}{(E_\nu/0.4~{\rm GeV})}^{2-s}\nonumber\\
&\times&D_{*,-2}^{7/10}M_{\rm ej,1}^{-21/20} {\mathcal E}_{\rm ej,51}^{27/20}t_{5.5}^{7/10}{\mathcal R}_{\rm cr10,1}^{-1}{(d/10~{\rm kpc})}^{-2},\,\,\,\,\,\,\,\,\,\,
\end{eqnarray}
where the factor $1/2$ comes from the facts that the $\pi^\pm/\pi^0$ ratio is $\approx2$ in $pp$ interactions and neutrinos carry $3/4$ of the pion energy in the decay chain. 
Also, ${\mathcal E}_s\approx L_st_{\rm dyn}$ and ${\mathcal R}_{\rm cr10}\equiv\epsilon_{\rm cr}\mathcal{E}_s/(E_p^2 dN_{\rm cr}/dE_p)|_{10~{\rm GeV}}$ is a spectrum dependent factor that converts the bolometric CR energy to the differential CR energy (e.g., Refs.~\cite{Murase:2013kda}). 

In Fig.~\ref{fig2}, we show neutrino energy fluences, evaluated at the time when the ratio of the signal to the square root of the background is a maximum (see below). The results agree with our analytical estimates, but the time window is chosen from the results of our detailed time-dependent model. Note that, for SNe II-P and IIn with an enhanced CSM, we expect ${\rm min}[1,f_{pp}]\sim1$ so the system is nearly ``calorimetric'' until some time.

\begin{table}[t]
\begin{center}
\caption{Expected numbers of through-going muon tracks in IceCube, for various types of SNe with different values of the muon energy threshold and observation time. 
\label{tb2}
}
\scalebox{0.95}{
\begin{tabular}{|c|c||c|c|c|c|c|}
\hline Model & $s$ & ${\mathcal N}_{\mu,>1~{\rm TeV}}^{\rm sig,<{10}^{7}~{\rm s}}$ & ${\mathcal N}_{\mu,>0.1~{\rm TeV}}^{\rm sig,<{10}^{7}~{\rm s}}$ &  ${\mathcal N}_{\mu,>0.1~{\rm TeV}}^{{\rm sig},<t_{\rm max}}$ & $t_{\rm max}$ [s] \\
\hline IIn & $2.2$ & $2.7\times{10}^{4}$ & $4.6\times{10}^{4}$ & $1.2\times{10}^{5}$ & $10^{7.5}$\\
($10$~kpc) & $2.0$ & $1.1\times{10}^{5}$ & $1.7\times{10}^{5}$ & $4.5\times{10}^{5}$ & ${10}^{7.5}$\\
\hline II-P\footnote{$D_*={10}^{-2}$ based on the observations of SN 2013fs (II-P).} & $2.2$ & $2.8\times{10}^2$ & $4.1\times{10}^2$  & $3.8\times{10}^2$ & ${10}^{5.8}$\\
($10$~kpc) & $2.0$ & $1.2\times{10}^3$ & $1.6\times{10}^3$ & $1.5\times{10}^3$ & ${10}^{5.8}$\\
\hline II-P\footnote{$D_*=1.34\times{10}^{-4}$ based on the observations of Betelgeuse (RSG).} & $2.2$ & $5.5\times{10}^2$ & $8.4\times{10}^2$ & $3.5\times{10}^2$ & ${10}^{5.4}$\\
($0.197$~kpc) & $2.0$ & $2.3\times{10}^3$ & $3.3\times{10}^3$ & $1.4\times{10}^3$ & ${10}^{5.4}$\\
\hline II-L/IIb & $2.2$ & $18$ & $27$ & $8.9$ & ${10}^{4.6}$\\
($10$~kpc) & $2.0$ & $78$ & $110$ & $36$ & ${10}^{4.6}$\\
\hline Ibc & $2.2$ & $5.4\times{10}^{-3}$ & $8.1\times{10}^{-3}$ & $2.8\times{10}^{-3}$ & ${10}^{3.8}$\\
($10$~kpc) & $2.0$ & $2.4\times{10}^{-2}$ & $3.2\times{10}^{-2}$ & $1.4\times{10}^{-2}$ & ${10}^{4.0}$\\
\hline
\end{tabular}
}
\end{center}
\end{table} 

\section{Detectability}
Taking into account neutrino mixing ($\nu_e:\nu_\mu:\nu_\tau\approx1:1:1$) and the IceCube angular resolution, we calculate the number of through-going muon tracks expected in IceCube (see Appendix of Ref.~\cite{Murase:2016gly}). 
The background increases as an integration time $t$, so the signal-to-background changes as time.  The time-dependent model is critical and our model is directly applicable to dedicated searches with IceCube and KM3Net.

\begin{figure}[tb]
\includegraphics[width=\linewidth]{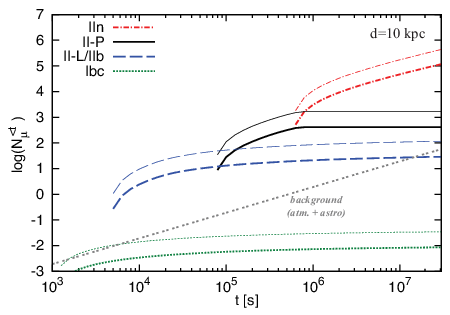}
\caption{Expected numbers of through-going muon tracks detected in IceCube, for a Galactic SN. For the signals, corresponding to Fig.~\ref{fig2}, we consider $s=2.2$ (thick curves) and $s=2.0$ (thin curves), respectively. The sum of atmospheric and astrophysical background events is also shown.
\label{fig3}
}
\end{figure}

In Fig.~\ref{fig3}, we show the number of muons expected in IceCube, ${\mathcal N}_{\mu,>E_{\rm th}}^{<t}$, which is integrated over muon energy ($E_\mu$) above a muon-energy threshold. 
We use IceCube's angle averaged effective area for upgoing tracks, which is sufficient for the purpose of this work. Note that KM3Net is more powerful for SNe in the southern sky, especially around the Galactic center.   
In Table~\ref{tb2}, we present the results for $t=10^7$~s, with $E_\mu^{\rm th}=0.1$~TeV and $E_\mu^{\rm th}=1$~TeV.  
We also calculate a test statistic, ${\mathcal N}_{\mu}^{\rm sig}/\sqrt{{\mathcal N}_{\mu}^{\rm bkg}}$, assuming both of the atmospheric (conventional+prompt) and astrophysical neutrino backgrounds~\cite{Aartsen:2016xlq}, which has a maximum at $t_{\rm max}$. For $w=2$, the test statistic increases as time as long as $f_{pp}\gtrsim1$, and then declines. Thus, $t_{\rm max}$ is essentially determined by $f_{pp}=1$ or $R_s=R_w$. From Fig.~\ref{fig3}, we see ${\mathcal N}_{\mu}^{\rm bkg}\lesssim1$ until $t\sim10^6$~s. For $E_\mu^{\rm th}=1$~TeV, the background is negligible in the relevant time window, because the atmospheric neutrino flux decreases as $\phi_{\nu}^{\rm atm}\propto E_\nu^{-3.7}$.   

We find that IceCube can detect $\sim10^3$ events for a Type II-P SN at 10~kpc. The core-collapse SN rate in the Milky Way is estimated to be $\sim3~{\rm century}^{-1}$~\cite{Adams:2013ana}, and the rates of Type II-P and II-L/IIb SNe are $\sim50$\% and $\sim15-20$\% of the core-collapse SNe, respectively~\cite{Smith:2010vz,Li:2010kc}. The frequency of rare SNe IIn is only $\sim10$\%~\cite{Margutti:2016wyh,Smith:2010vz}, but they can be detected up to $\sim10$~Mpc. 
Type Ibc SNe, whose rate is $\sim20-25$\%~\cite{Smith:2010vz,Shivvers:2016bnc}, may be difficult to detect. However, they are also detectable if a dense CSM is expelled as suggested for low-luminosity $\gamma$-ray bursts~\cite{Kashiyama:2012zn}. While the rate is much smaller, nearby SNe are more spectacular. For Betelgeuse, we expect $\sim3\times{10}^6$ ($\sim{10}^3$) with (without) an enhanced CSM. Another example is $\eta$ Carinae at 2.3~kpc. If it explodes as a SN IIn (assuming that a CSM with $w=2$ is erupted), we could have $\sim3\times10^6$ neutrino events~\footnote{The observed nebula~\cite{Smith:2012nw} implies that $\sim10^5$ neutrinos can be seen if $\eta$ Carinae explodes without an additional CSM~\cite{Murase:2010cu}, but more detailed modeling is necessary.}. 

\section{Summary and Implications}
For the first time, we systematically evaluate the detectability of high-energy neutrinos from different classes of nearby SNe in a common and well-justified setup.
For a Galactic SN, surprisingly, we showed that Gton neutrino detectors like IceCube and KM3Net can detect $\gtrsim10$ neutrinos from a Type II SN, which forms a dominant class of core-collapse SNe. Even $\sim10^3-10^6$ events are possible for Type II-P and IIn SNe. The targeted time window is $\sim0.1-1$~day ($\sim0.1-1$~yr) after the MeV neutrino detection for SNe II-P/II-L/IIb (SNe IIn). 
Since the nearest SNe are intermittent, we must be best prepared for them. Exploiting the time-dependent model is crucial for not missing the neutrino signals. The signal-to-background changes as time, and dedicated searches optimized in both time and energy can be achieved with our new model. In the simplest model, the expected signal can be significantly underestimated, and the detectability is diminished. 
This is especially critical for detections of a SN with a smaller CSM or in another galaxy (e.g., M31) and stacking analyses with multiple SNe. The SN rate within 10 Mpc is enhanced due to a local overdensity and a higher-than-expected SN rate within 10 Mpc~\cite{Horiuchi:2011zz,Nakamura:2016kkl}.

In our model, CR production relies on the shock acceleration, and target material is not the SN ejecta but the ``CSM''.
We stress that our new prediction qualitatively differs from others, e.g., one from interactions between pulsar-accelerated CRs and the ejecta~\cite{Gaisser:1987wm,Murase:2009pg,Fang:2013vla}, and includes ``dominant'' SNe II-P, unlike previous works for rarer SNe IIn and Ibc-BL~\citep{Murase:2010cu,Katz:2011zx,Kashiyama:2012zn}. It is clearly consistent with the nonobservations from past SNe such as SN 1987A, but luminous SNe IIn in nearby galaxies~\cite{Horiuchi:2011zz,Nakamura:2016kkl} could be seen as ``mini-flares''. 
Note that our model is also consistent with the existing $\gamma$-ray constraints on SNe IIn~\cite{TheFermiLAT:2015kla}. The Type IIn contribution to the diffuse neutrino intensity could reach $E_\nu^2\Phi_\nu\sim3\times{10}^{-8}~{\rm GeV}~{\rm cm}^{-2}~{\rm s}^{-1}~{\rm sr}^{-1}$, whereas Eq.~(7) suggests that the Type II-P contribution is $\sim1$\% of that.  
 
There are various implications. 
(a) First, we can probe mechanisms of the pre-SN mass ejection or envelope inflation. 
Unlike photons that can be largely attenuated, neutrinos give us direct information on the CSM. 
(b) Second, through the observation of $\sim0.1-1$~PeV neutrinos, we can test whether nascent SN remnants contribute to the observed CRs around the knee at $3$~PeV. This is relevant, since SNe IIn and IIb have been suggested as the origins of such very high-energy CRs~\cite{Murase:2013kda,Sveshnikova:2003sa,Ptuskin:2010zn}.  
(c) Third, we can study CR ion acceleration in ``real time''.  Neutrino signals earlier than $t_{\rm onset}$ may indicate CR acceleration inside the ejecta~\cite{Murase:2013mpa,Murase:2009pg,Fang:2013vla,Waxman:2001kt}.
(d) The Galactic SN is an ideal target for multimessenger astrophysics. For example, thermal radiation in the optical and x-ray bands will coincide with high-energy neutrino emission. 
We also predict the generation of hadronic $\gamma$ rays from the production and decay of neutral pions. If the system is transparent to $\gamma$ rays, the Galactic SNe should readily be detected by current $\gamma$-ray telescopes, {\it Fermi}, HAWC and the future Cherenkov Telescope Array. In particular, {\it Fermi} could see SNe II-P up to $\sim1-2$~Mpc. 

Regarding (b), the highest-energy neutrinos can be seen in different flavors. Based on the published IceCube effective areas~\cite{Aartsen:2013jdh,Aartsen:2015dlt}, we may expect a few double-bang (or double-pulse) events and even Glashow resonance events for a Galactic SN II-P. Statistics can be improved with a future detector such as {\it IceCube-Gen2}~\cite{Aartsen:2014njl}, and observations of $\nu_\tau$ and $\bar{\nu}_e$ will give us information on flavor mixing and neutrino production mechanisms. 

High-energy neutrino detections with high statistics will give us unique opportunities to study neutrino properties, e.g., neutrino decay~\cite{Beacom:2002vi,Baerwald:2012kc,Pagliaroli:2015rca,Shoemaker:2015qul}, oscillation into other sterile neutrinos~\cite{Beacom:2003eu,Pakvasa:2012db,Shoemaker:2015qul,Bustamante:2016ciw}, and neutrino self-interactions~\cite{Ioka:2014kca,Ng:2014pca,Ibe:2014pja,Araki:2014ona,Blum:2014ewa,Shoemaker:2015qul}. Note that the CSM is not dense enough for the matter resonance in the sources to occur at relevant energies. Flavor studies can be used to probe small mass-splittings of pseudo-Dirac neutrinos with ${\Delta m}_{j}^2\approx8\times{10}^{-15}~{({\rm eV}/c^2)}^2~(E_\nu/1~{\rm TeV}){(d/10~{\rm kpc})}^{-1}$. 
A bright neutrino point source with a long duration will also enable us to do the Earth tomography, and provide a new test of the cross section. 

Related (c), the detection of GeV-TeV neutrinos is relevant to obtain broadband spectra, which is possible with Hyper-Kamiokande (Hyper-K)~\cite{Abe:2011ts}, KM3Net-ORCA~\cite{Adrian-Martinez:2016fdl}, and PINGU~\cite{Koskinen:2011zz}. 
By detecting nonthermal neutrinos from GeV to PeV energies, as well as thermal MeV neutrinos, the next Galactic SN will provide the first example of ``multienergy'' neutrino astrophysics. 


\medskip
\begin{acknowledgments}
K.M. thanks John Beacom for his helpful comments. K.M. also thanks Markus Ahlers, Roger Chevalier, Doug Cowen, Francis Halzen, Shunsaku Horiuchi, Marek Kowalski, Maria Petropoulou, Todd Thompson, Takatomi Yano, and Shigeru Yoshida for useful discussions. K.M. acknowledges the workshop, ``Cosmic Neutrino Observations with Hyper-Kamiokande'' in 2015, where preliminary estimates of this work were presented. The work of K.M. is supported by the Alfred P. Sloan Foundation and the U.S. National Science Foundation (NSF) under grant No. PHY-1620777.
\end{acknowledgments}

\bibliography{kmurase.bib}

\newpage
\appendix

\setcounter{equation}{0}
\setcounter{figure}{0}
\setcounter{table}{0}
\setcounter{section}{0}
\setcounter{page}{1}
\makeatletter
\renewcommand{\theequation}{S\arabic{equation}}
\renewcommand{\thefigure}{S\arabic{figure}}
\renewcommand{\thetable}{S\arabic{table}}
\newcommand\ptwiddle[1]{\mathord{\mathop{#1}\limits^{\scriptscriptstyle(\sim)}}}

\section{Supplemental Material}
Following the supernova (SN) dynamics, we assume a power-law cosmic-ray (CR) spectrum, 
\begin{equation}
\frac{dn_{{\rm cr}}}{dp}\propto{p}^{-s}e^{-p/p_{\rm max}}.
\end{equation}
Here the maximum momentum, $p_{\rm max}$, is set by the competition between the CR acceleration time and CR cooling time due to various energy loss processes such as inelastic $pp$ interactions, photohadronic interactions, and adiabatic losses. The normalization of the CR spectrum is set by the CR energy density,
\begin{equation}
U_{\rm cr}=\int^{p_{\rm max}} dp \,\,\, E_p \frac{dn_{{\rm cr}}}{dp},
\end{equation}
where $U_{\rm cr}\approx \epsilon_pL_s/(4\pi R_s^2V_s)$. We also parameterize the magnetic field in the emission region via $U_B\equiv\varepsilon_B{\mathcal E}_s/{\mathcal V}$, where $\mathcal V\approx(4\pi/3)R_s^3$ is the volume, and $\varepsilon_B=0.01$ is used in this work. Note that this definition is slightly different from that against the ram pressure.

We consistently calculate differential energy densities of neutrinos, gamma-rays, and electrons/positrons, in a time-dependent manner. 
The differential energy injection rates are calculated by the following formulas:
\begin{eqnarray}
\dot{n}_{E_\nu}^{\rm inj}=\frac{d\sigma_{pp}\xi_\nu}{dE_\nu} \frac{cM_{\rm cs}}{m_H\mathcal V} \int^{p_{\rm max}} dp \,\,\, \frac{dn_{{\rm cr}}}{dp}\nonumber\\
\dot{n}_{E_\gamma}^{\rm inj}=\frac{d\sigma_{pp}\xi_\gamma}{dE_\gamma} \frac{cM_{\rm cs}}{m_H\mathcal V}\int^{p_{\rm max}} dp \,\,\, \frac{dn_{{\rm cr}}}{dp}\nonumber\\
\dot{n}_{E_e}^{\rm inj}=\frac{d\sigma_{pp}\xi_e}{dE_e} \frac{cM_{\rm cs}}{m_H\mathcal V}\int^{p_{\rm max}} dp \,\,\, \frac{dn_{{\rm cr}}}{dp},
\end{eqnarray}
where $\xi_\nu$, $\xi_\gamma$ and $\xi_e$ are multiplicities of neutrinos, gamma rays and electrons/positrons, respectively.  We calculate the differential spectra of secondary particles following the parametrization by Ref.~\cite{Kelner:2006tc}, and we use the total $pp$ cross section following the post-Large-Hadron-Collider formula given by Ref.~\cite{Kafexhiu:2014cua}.  

The differential neutrino luminosity is calculated by 
\begin{equation}
E_\nu L_{E_\nu}= \frac{(E_\nu^2 n_{E_\nu}){\mathcal V}}{t_{\rm esc}},
\end{equation}
where $t_{\rm esc}\approx R_s/c$. 
Very importantly, our time-dependent model contains only a few free parameters, $\epsilon_p$, $s$, and $\varepsilon_B$. All the parameters that are related to the SN shock dynamics can be determined by the observations and modeling of optical, x-ray, and radio emissions from the SN. 
This is likely to happen for the nearby SNe since detailed electromagnetic observations will be available. 

The method used in this work allows us to consistently evaluate gamma-ray emission in a simultaneous manner. Details are rather complicated and will be presented elsewhere. We solve the following kinetic equations: 
\begin{eqnarray}\label{eq:cascade}
\dot{n}_{E_e}^e &=& \dot{n}_{E_e}^{(\gamma\gamma)}
- \frac{\pd}{\pd E_e} [(P_{\rm IC}+P_{\rm syn}+P_{\rm ad}+P_{\rm Cou}) n_{E_e}^e] + \dot{n}_{E_e}^{\rm inj}\nonumber\\ 
\dot{n}_{E_\gamma}^\gamma &=& -\frac{n_{E_\gamma}^{\gamma}}{t_{\gamma \gamma}} -\frac{n_{E_\gamma}^{\gamma}}{t_{\rm matter}} - \frac{n_{E_\gamma}^{\gamma}}{t_{\rm esc}}
+ \frac{\pd n_{E_\gamma}^{(\rm IC)}}{\pd t} 
+ \frac{\pd n_{E_\gamma}^{(\rm syn)}}{\pd t} + \dot{n}_{E_\gamma}^{\rm inj}\nonumber\\ 
\end{eqnarray}
where 
\begin{eqnarray}
t_{\gamma \gamma}^{-1} &=& \int d E_\gamma \,\, n_{E_\gamma}^\gamma  \int \frac{d \cos\theta}{2} \,\, \tilde{c} \sigma_{\gamma \gamma} 
, \nonumber \\
\dot{n}_{E_\gamma}^{(\rm IC)} &=& \int d E_e \,\, n_{E_e}^e \, \int d E_\gamma\,\, n_{E_\gamma}^\gamma \, \int  \frac{d \cos\theta}{2} \,\, \tilde{c}   \frac{d \sigma_{\rm IC}}{d E_\gamma} 
, \nonumber \\
\dot{n}_{E_e}^{(\gamma \gamma)} &=& \frac{1}{2} \int d E_\gamma\,\, n_{E_\gamma}^\gamma \, \int d E'_\gamma\,\, n_{E'_\gamma}^\gamma \, \int \frac{d \cos\theta}{2} \,\, \tilde{c}  \frac{d \sigma_{\gamma \gamma}}{d E_e} 
, \nonumber
\end{eqnarray}
where $\tilde{c} =(1-\cos\theta)c$ (where $\theta$ is the angle between two particles), $t_{\gamma\gamma}$ is the two-photon annihilation time, $t_{\rm matter}$ is the energy loss time scale due to interactions with matter, $t_{\rm esc}\approx R_s/c$ is the photon escape time, $P_{\rm IC}$, $P_{\rm syn}$, $P_{\rm ad}$, and $P_{\rm Cou}$ are energy loss rates for the inverse-Compton radiation, synchrotron radiation, adiabatic cooling, and Coulomb collisions, respectively. The calculation is performed during the dynamical time $t_{\rm dyn}\approx R_s/V_s$. Note that the method used here takes full account of electromagnetic cascades, unlike  the previous work~\cite{Murase:2010cu}.

We also consider SN radiation fields themselves, although this does not influence the results on neutrino spectra. There are two components. The first component is a thermal bremsstrahlung component, and the other is an SN optical emission component. The former is implemented based on the shock temperature and the post-shock density, although the x-ray photons do not affect the results of this work. For the latter, the radiation energy density is estimated to be
\begin{equation}
U_{\rm sn}\approx\frac{(1+\tau_T)3L_{\rm sn}}{4\pi R_s^2 c},
\end{equation}
where $\tau_T$ is the Thomson optical depth. The SN optical luminosity is given by $L_{\rm sn}=\epsilon_{\rm sn}L_{s}$, where $\epsilon_{\rm sn}=1/4$ is assumed. We also use a gray body spectrum, assuming the SN temperature to be ${\mathcal T}_{\rm sn}\approx\rm max[{\mathcal T}_{\rm bb},{\mathcal T}_{\rm rec}]$, where ${\mathcal T}_{\rm bb}$ is the black body temperature and ${\mathcal T}_{\rm rec}=10^4$~K. 


\end{document}